\input{aipcheck}

\documentclass[
  ,draft            
  ]
  {aipproc}

\layoutstyle{6x9}

\begin{document}

\title[Feeding and Feedback in NGC 4151]{Feeding and Feedback in NGC 4151 from GEMINI Near Infrared Integral Field Spectroscopy}


\author{Ramiro D. Sim\~oes Lopes}{
address={Instituto de F\'isica, UFRGS, Av. Bento Gon\c{c}alves 9500, 91501-970, Porto Alegre, RS, Brazil},
email={ramirosl@if.ufrgs.br}
}

\author{Thaisa Storchi-Bergmann}{
address={Instituto de F\'isica, UFRGS, Av. Bento Gon\c{c}alves 9500, 91501-970, Porto Alegre, RS, Brazil},}

\author{Rogemar A. Riffel}{
address={Instituto de F\'isica, UFRGS, Av. Bento Gon\c{c}alves 9500, 91501-970, Porto Alegre, RS, Brazil},}

\author{Peter J. McGregor}{
address={Research School of Astronomy and Astrophysics, ANU, Weston Creek, ACT2611, Australia}
}

\author{Paul Martini}{
address={Department of Astronomy and Center for Cosmology and Astroparticle Physics, The Ohio State University, Columbus, OH, USA}
}

\begin{abstract}
We discuss two-dimensional mapping of the near-infrared  emission-line intensity distributions and kinematics of the narrow-line region (NLR) of NGC\,4151, obtained with the Gemini Near-Infrared Integral Field Spectrograph, with a projected spatial resolution of $\approx$\,8\,pc. The ionized gas intensity distribution follows the projected bi-cone morphology observed in  previous optical narrow-band images, and its kinematics reveal outflows along the bi-cone. We propose a kinematic model in which the gas in the NLR moves at a velocity of $\approx\,600$\,km\,s$^{-1}$ up to $\sim$\,100\,pc from the nucleus. A completely distinct morphology and kinematic structure is observed for the molecular gas, which avoids the region of the bi-cone and has velocities close to systemic, and is consistent with an origin in the galaxy plane. The molecular gas thus traces the AGN feeding, while the ionized gas traces its feedback.
\end{abstract}

\maketitle
\section{Introduction}

NGC\,4151 is the nearest (13.3\,Mpc) and apparently brightest Seyfert~1 galaxy, harboring one of the best studied active galactic nuclei (hereafter AGN). In the optical, the emitting gas of the  narrow-line region (hereafter NLR) has  the morphology of a projected bi-cone, as observed in a previous Hubble Space Telescope (HST) [O\,{\sc iii}]$\lambda$5007 narrow-band image. According to previous studies, our line-of-sight is outside but close to the edge of the cones \citep{evans93,hutchings98,das05}, which are oriented along  position angle (PA) $\approx60^\circ$. Optical long-slit spectroscopy reveals outflows along the cones, with the approaching side to the SW. In the radio there is a linear structure along PA$=77^\circ$ that is not aligned with the bi-cone \citep{mundell95}.  We have observed the inner $3\times8$\,arcsec$^2$ of NGC\,4151 on the Gemini North telescope with the Near-Infrared Integral Field Spectrograph ( NIFS) \citep{mcgregor03} operating with the ALTAIR adaptive optics module in order to map the NLR gas excitation and kinematics with complete two-dimensional coverage in the near-infrared, a spectral region less affected by obscuration than the optical. The spectral range covered was 0.94-2.51\,$\mu$m at two-pixel resolving power  $\approx$5200 and angular resolution of 0.12 arcsec, corresponding to $\approx$\,8 pc at the galaxy. 

\begin{figure}[t]
 \centering
 \begin{minipage}{0.5\linewidth}
 \includegraphics[height=.5\textheight]{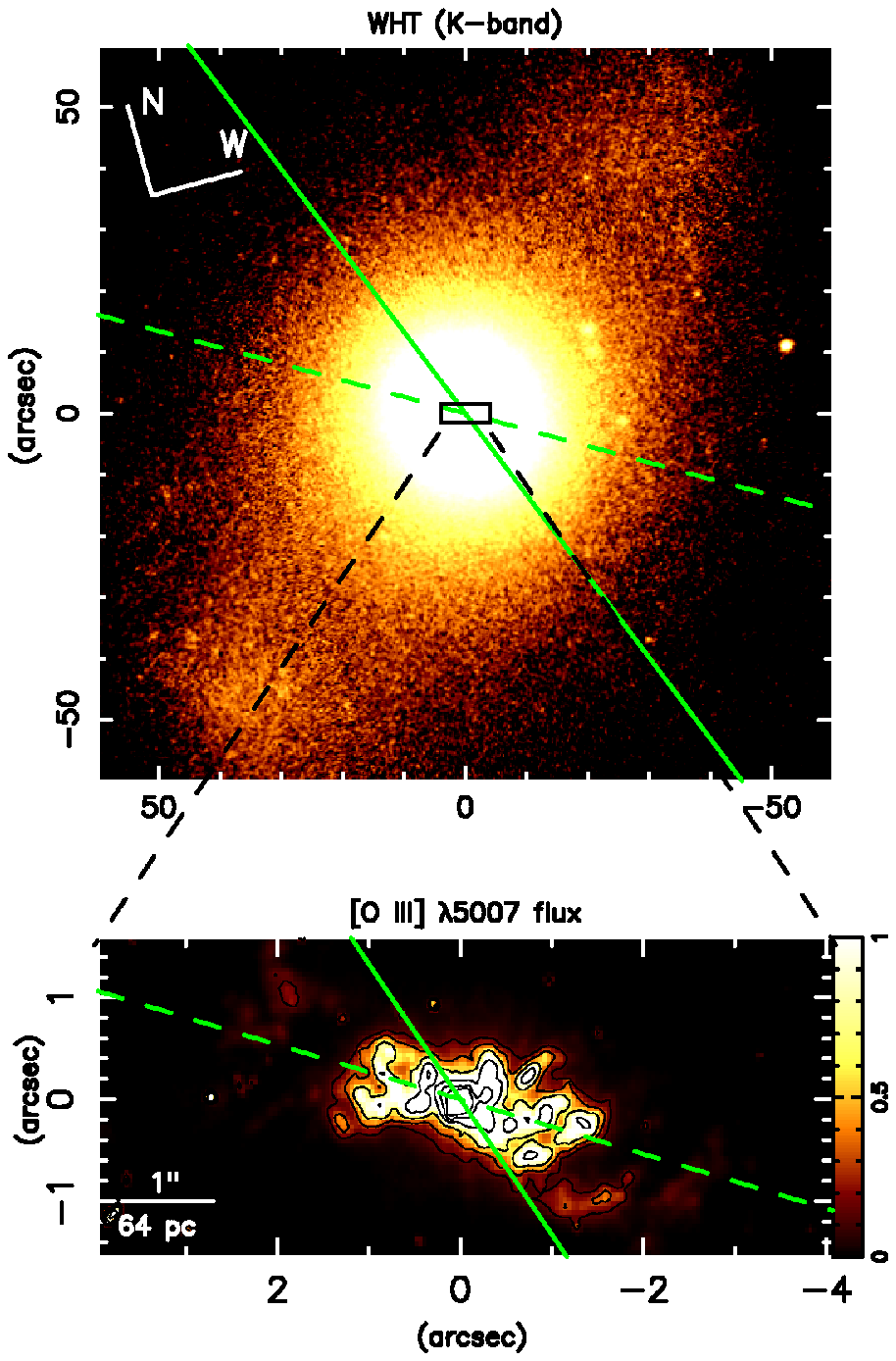}
 \end{minipage}
 \begin{minipage}{0.5\linewidth}
 \includegraphics[height=.5\textheight]{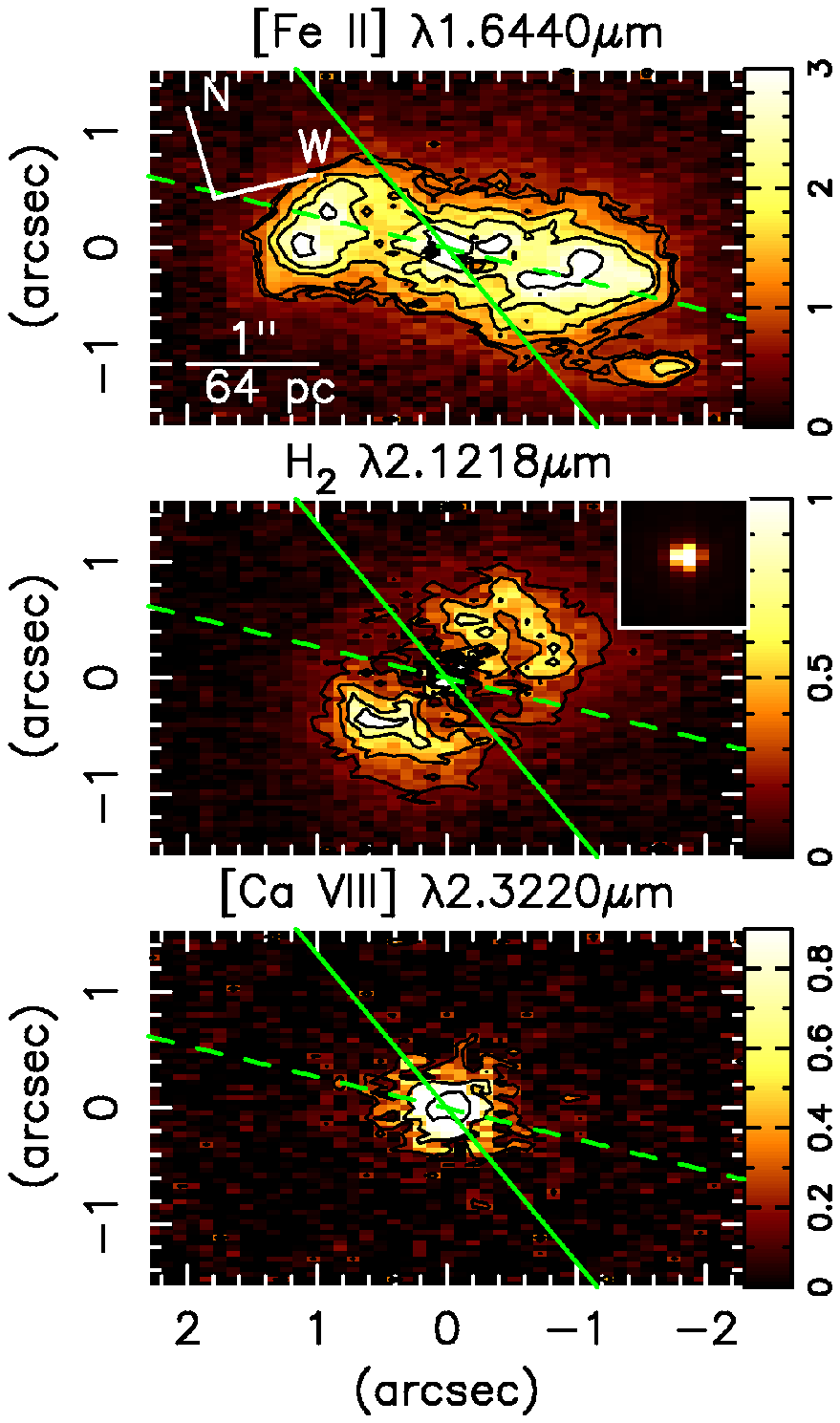}
 \end{minipage}
 \caption{LEFT -- Top: K-band image of the central 60$^\circ\times$60$^\circ$ of NGC\,4151 obtained with the William Herschel Telescope. The continuous line shows the orientation of the major axis of the galaxy, while the dashed line shows the orientation of the bi-cone. The rectangle shows the region covered by the NIFS observations. Bottom: HST [O\,{\sc iii}]$\lambda$5007 narrow-band image of the NLR in the field-of-view of the NIFS observations. RIGHT --  Intensity maps in the inner 3$\times$5\,arcsec$^2$.Top:  [Fe\,{\sc ii}] intensity map;  middle: H$_2$ intensity map;  bottom: a coronal line intensity map. }
 \label{large}  
 \end{figure}

In the top-left panel of Fig.~\ref{large}, we present a K-band image of the central $60\times60$~arcsec$^2$ of NGC\,4151, where the large-scale bar can  be observed at PA=130$^\circ$. The position of the major axis of the galaxy (22$^\circ$) is almost perpendicular to the bar. The central rectangle shows the field of view covered by the NIFS observations. In the bottom-left panel we show the HST [O\,{\sc iii}]$\lambda$5007 narrow-band image of the NLR in the field-of-view of the NIFS observations.



\section{NLR Intensity Distributions and Excitation}

We have maped the intensity distributions in 14 emission lines. Three distinct distributions are observed, as illustrated in Fig.\,\ref{large}. Most of the ionized gas is extended to  $\approx$100~pc from the nucleus along the region covered by  the known bi-conical outflow,  as observed in the [Fe\,{\sc ii}]  intensity map (top right panel of Fig.\,\ref{large}). The molecular gas  (H$_2$) avoids the region of the bi-cone, extending from $\approx$10  to 60\,pc  from the nucleus approximately along the large scale bar, almost perpendicular to the bi-cone axis (middle right panel of  Fig.\,\ref{large}). The molecular gas appears to be destroyed by the intense nuclear radiation 
escaping along the bi-cone, and its intensity distribution suggests an origin in the galaxy plane. The coronal line region is only barely resolved (bottom right panel of  Fig.\,\ref{large}), consistent with an origin in the inner NLR.

The line ratios [Fe{\, \sc ii}]1.257$\mu$m/[P\,{\sc ii}]1.189$\mu$m  and [Fe{\, \sc ii}]1.257$\mu$m/Pa$\beta$ 
of the NLR of NGC\,4151  correlate with the radio intensity distribution, mapping the effects of shocks produced by the radio jet on the NLR. These shocks probably release the Fe locked in grains and  produce the observed enhancement of the [Fe{\, \sc ii}] emission at  $\approx$1 arcsec from the nucleus (top right panel of  Fig.\,\ref{large}). In these regions, we obtain electron densities  $n_e\approx4000\,{\rm cm^{-3}}$ and temperatures $T_e \approx 15000$\,K for the [Fe{\, \sc ii}]-emitting gas. For the H$_2$-emitting gas,  we obtain much lower temperatures of $T_{\rm exc} \approx 2100$\,K and conclude that the gas is in thermal equilibrium. The heating necessary to excite the molecule may be due to X-rays escaping through the nuclear torus \citep{riffel09}. The escape of radiation through the torus is also required by the observation that the bi-cone does not have a sharp apex, but rather show ionized gas emission in all directions around the nucleus.

\section{Gas Kinematics}

\begin{figure}
  \includegraphics[scale=0.7, angle=-90]{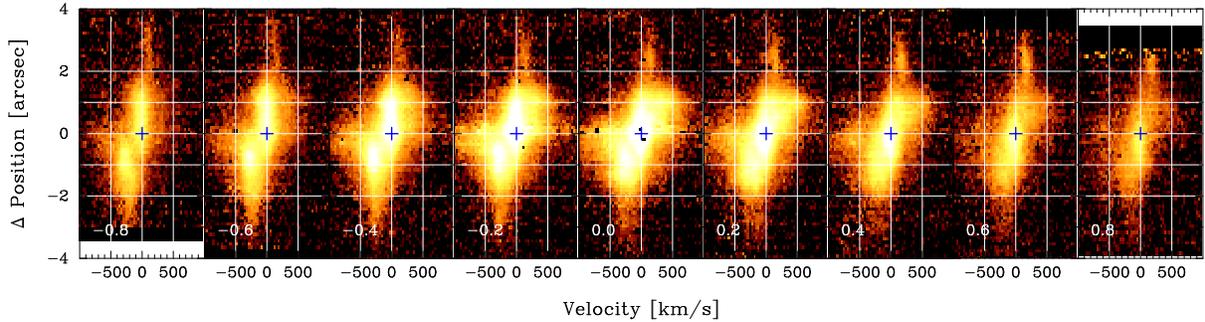}
  \caption{Position-velocity diagrams of [S\,{\sc iii}] 0.9533\,$\mu$m emission for 0.2 arcsec wide pseudo-slits oriented at PA=60$^\circ$ along the bi-cone axis and offset perpendicular to the pseudo-slit by $-$0.8 arcsec to $+$0.8 arcsec as indicated in each frame. An average continuum image extracted between $\pm(2000-2500)$~km\,s$^{-1}$ has been subtracted. Crosses mark the location of the nucleus in the central pseudo-slit and the ``equivalent'' location for the other slits.}
\label{pv}
\end{figure}

\begin{figure}
  \includegraphics[scale=0.6, angle=-90]{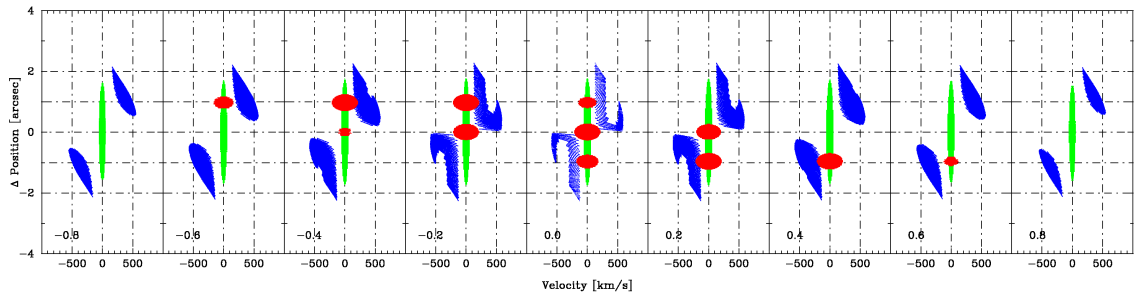}
  \caption{Simulated position-velocity diagrams for a truncated hollow cone model with constant velocity of 600~km\,s$^{-1}$ outflow. Besides the bi-conical outflow zero velocity components are shown for the emission associated with the radio jet (the three knots) and for the galactic disk component (sequence at zero velocity).}
\label{pv-mod}
\end{figure}

In Fig.\,\ref{pv}, we present a position-velocity (PV) diagram for the [S\,{\sc iii}] emission line along the bi-cone axis. We modelled the NLR kinematics as a truncated hollow conical outflow at a constant velocity of 600\,km\,s$^{-1}$. We did not see acceleration or deceleration in our data, as proposed by \citet{das05} from the modeling of the [O\,{\sc iii}] emitting-gas kinematics. The model PV diagram is shown in Fig.\,\ref{pv-mod}. Besides the outflowing component, the ionized gas shows two components at velocities close to systemic. One is gas from the galaxy plane, ionized by  radiation escaping through the torus, as discussed above, and the origin of the sequences at zero velocity  in Fig.\,\ref{pv-mod}. The other is emission at zero velocity generated by the interaction between the radio jet  and gas from the galaxy plane (the three knots in Fig.\,\ref{pv-mod}).

We estimated a mass outflow rate of $\approx{\rm 1 M_\odot\,yr^{-1}}$ along each cone, which exceeds the inferred black hole accretion  rate by a factor of $\approx$\,100. This can be understood if the NLR is formed mostly by entrained gas from the circumnuclear interstellar medium by an outflow that probably originates in the accretion disk.

The H$_2$ emission, besides presenting a distinct intensity  distribution, also has a distinct kinematic structure from the ionized gas. The velocities are close to systemic and consistent with an origin  in the galaxy plane. This hot molecular gas may only be the tracer of a larger reservoir of colder gas which represents the AGN feeding, while the ionized gas outflow represents the AGN feedback.

More details on the NLR excitation and kinematics can be found in \citet{thaisa09a} and \citet{thaisa09b}, 
respectively. The analysis of the nuclear continuum -- which reveals the presence of an unresolved infrared source that can be attributed to a dusty torus ---  is discussed in \citet{riffel09}.





\bibliographystyle{aipproc}   

\bibliography{simoeslopes_r}

\IfFileExists{\jobname.bbl}{}
 {\typeout{}
  \typeout{******************************************}
  \typeout{** Please run "bibtex \jobname" to optain}
  \typeout{** the bibliography and then re-run LaTeX}
  \typeout{** twice to fix the references!}
  \typeout{******************************************}
  \typeout{}
 }


\end{document}